\documentclass[3p]{elsarticle}

\usepackage{amsmath}
\usepackage{amssymb}

\newcommand{\beq}{\begin{equation}}
\newcommand{\eeq}{\end{equation}}

\newcommand{\curl}{\nabla\times}
\newcommand{\vv}{\mathbf{v}}

\newcommand{\lap}{\Delta}
\newcommand{\ilap}{\Delta^{-1}}

\newcommand{\hz}{\hat{\mathbf{z}}}
\newcommand{\bp}{{\bar{\phi}}}
\newcommand{\tp}{{\tilde{\phi}}}
\newcommand{\ben}{\begin{eqnarray}}
\newcommand{\een}{\end{eqnarray}}
\newcommand{\benn}{\begin{eqnarray*}}
\newcommand{\eenn}{\end{eqnarray*}}

\newcommand{\calL}{{\mathcal L}}
\newcommand{\hN}{\hat{N}}

\begin{document}
\begin{frontmatter}

\title{Hamiltonian formulation of the modified Hasegawa Mima equation}

\author{C. Chandre$^{1,2}$, P.J. Morrison$^3$,  E. Tassi$^{1,2}$}
\address{$^1$ Aix-Marseille Universit\'e, CNRS, CPT, UMR 7332, 13288 Marseille, France \\
$^2$ Universit\'e de Toulon, CNRS, CPT, UMR 7332, 83957 La Garde, France\\
$^3$ Institute for Fusion Studies and Department of Physics, The University of Texas at Austin, Austin, TX 78712-1060, USA }
%\date{\today}

\begin{abstract}
We derive the Hamiltonian structure of the modified Hasegawa-Mima equation from the ion fluid equations applying Dirac's theory of constraints. We discuss the Casimirs obtained from the corresponding Poisson structure.  
\end{abstract}
\end{frontmatter}

%%%%%%%%%%%%%%%%%%%%%%%%%%

%%%%%%%%%%%%%%%%%%%%%%%%%%
%\section{Introduction}

Zonal flows are believed to have a dramatic effect on the confinement of magnetized plasmas by suppressing the associated turbulent transport, notably the radial transport.   Through the years various reduced  models have been developed for describing transport in toroidal plasmas,   and  zonal flows have been observed in simulations of some of these models.  An early example of such a reduced model is the Hasegawa-Mima (HM) equation \cite{HM}, but more general  electrostatic fluid models using both the FLR  (e.g., \cite{HH})  and gyrofluid (e.g., \cite{HKM,HDP})  approaches have been available for many years.  Similarly, electromagnetic gyrofluid models (e.g., \cite{SH,WHM}) have been developed.  Some of these models  possess noncanonical Hamiltonian structure (see \cite{morr82,morrison98,morr05}), which has been used to guide  construction and led to the identification of new and physically important terms (e.g., \cite{HHM}), and  has also been shown to be important for the consistent calculation of zonal flow dynamics (e.g., \cite{KK1,KK3}). 

Zonal flows are also believed to be a nonlinear manifestation of  drift wave modulational instability, the physics of which is contained in the HM equation.  However, it  was recognized in  \cite{dorland} that this physics is more accurately described by a modified form of the HM equation for which the adiabatic electron response  is modified so as to take into account the geometry of magnetic surfaces.  For example, it is shown in \cite{dewar} that the modification  enhances the generation of zonal flows.  
 
The purpose of the present contribution is to demonstrate that the modified Hasegawa-Mima (mHM) equation possesses Hamiltonian structure by obtaining if  from Dirac's theory of constrained Hamiltonian systems \cite{dira50,suda74,Bhans76,sund82}, a technique  used in previous derivations \cite{morr09,chandre,chandre2}.  It is known that the modification of the HM equation applies for more general multifield theories (e.g., \cite{WMH});  consequently,  the  methods we use and the results we obtain are of general utility and can be adapted to apply to a very large class of  reduced fluid models. 

For simplicity of our argumentation, we consider a slab geometry in which $x$ corresponds to the radial direction and $y$ to the poloidal angle. The zonal part of any field $\chi(x,y)$ is given by $\tilde{\chi}=\chi-\bar{\chi}$,  
where 
$$
\bar{\chi}=\frac{1}{2\pi}\int_0^{2\pi}\!\!dy\,  \chi \,.
$$
First we consider a parent model with two dynamical equations from which the mHM equation can be derived: one describing the transverse dynamics of the ion velocity field $\vv(x,y,t)$ and the other describing the dynamics of the ion density field $n(x,y,t)$:
\ben
&& \dot{\vv} +(\vv \cdot \nabla) \vv=-\nabla\varphi+\vv \times {\bf B},\label{eqn:s1}\\
&& \dot{n}=-\nabla\cdot (n\vv),\label{eqn:s2}
\een
where the dot indicates the partial derivative with respect to time $t$. Here we use units such that the ion mass is $M=1$, its charge $e=1$, and the amplitude of the magnetic field $B=1$. The usual HM derivation is by a rather straightforward combination of the two equations for the density and the velocity field, assuming that the ion polarization velocity is much smaller than the ${\bf E}\times{\bf B}$ drift. In previous work \cite{chandre} we have  shown a different way of deriving this equation from the Hamiltonian structure of the parent model. This method  of derivation allows one to derive the reduced equation with the Hamiltonian structure  naturally provided.   

The total energy of the ions, given by the sum of their kinetic energy plus the potential energy provided by the electrostatic potential $\varphi$, is a conserved quantity that is also  the Hamiltonian of the system of Eqs.~(\ref{eqn:s1} and \ref{eqn:s2}), viz.
\beq
H[n,\mathbf{v}]=\int d^2 x \left[n\frac{v^2}{2}+n\varphi\right].
\label{psiext}
\eeq
The dynamics is determined by the Poisson bracket~\cite{MG80,tass09}
\beq \label{eq:PBparent}
\{F,G\}=-\int d^2 x \left[F_{\mathbf{v}}\cdot \nabla G_n - \nabla F_n\cdot G_{\mathbf{v}}-\left(\frac{\curl \bf{v}+\hat{\bf z}}{n}\right)\cdot F_{\mathbf{v}}\times G_{\mathbf{v}}\right],
\eeq
where we denote the functional derivatives of a given observable $F[n,\vv]$ by subscripts, i.e.\ $F_{\vv}=\delta F/\delta \vv$ and $F_n=\delta F/\delta n$.
In the present context we assume that the electrostatic potential $\varphi$ is determined by the dynamics of the electrons which leads to a function $\varphi(n_e)$, where $n_e$ is the electron density. From the quasi-neutrality condition, $n=n_e$,  the Hamiltonian becomes 
\beq
H[n,\mathbf{v}]=\int d^2 x \left[n\frac{v^2}{2}+\psi(n)\right],
 \label{eq:Hparent}
 \eeq
where $\psi'(n)=\varphi(n)$. The usual HM  equation is obtained by neglecting the inertia of the electrons so that their density obeys a Boltzmann law $n_e=n_0 \exp\varphi$, where $n_0=n_0(x,y)=1-\lambda(x,y)$ is the electron density at equilibrium. For the mHM equation, this adiabatic response has to take into account the prescription of \cite{dorland}, which here reads $n_e=n_0 \exp\tilde{\varphi}$,   where $\tilde{\varphi}$ is the zonal part of the potential.
 
Next we perform a change of variables $(n,\vv)\mapsto (n,\phi,D)$ where
\beq
 \lap \phi=\hat{\bf z}\cdot\nabla\times \vv,\qquad  \lap D=\nabla\cdot \vv\,,
\eeq
where $\lap$ denotes the Laplacian.
In terms of the new variables $(n,\phi,D)$, the Hamiltonian of (\ref{eq:Hparent}) becomes
\begin{equation}
\label{eq:Hion}
H[n,\phi,D]=\int d^2 x \left[n\left(\frac{|\nabla \phi|^2+|\nabla D|^2}{2}+[ \phi,D]\right)+\psi(n)\right],
\end{equation}
where $[f,g]=\hat{\bf z}\cdot \nabla f \times \nabla g$, and the bracket~(\ref{eq:PBparent}) becomes
\ben
\{F,G\}&=&\int\! d^2 x\,  \Big[ F_n G_D-F_DG_n-F_\phi\ilap \calL\ilap G_\phi-F_D\ilap\calL\ilap G_D
\nonumber\\
& & \hspace{1.5 cm} +F_D\ilap\Lambda\ilap G_\phi-F_\phi \ilap\Lambda\ilap G_D\Big],
\label{eq:PBion}
\een
where the two linear operators $\calL$ and $\Lambda$ acting on a function $f({\bf x})$ are defined by
\beq
 \calL f = \left[\frac{\Delta \phi+1}{n}, f  \right],\qquad 
 \Lambda f = -\nabla\cdot \left( \frac{\Delta \phi+1}{n} \nabla f\right).
\eeq

%\section{Modified Hasegawa-Mima equation}

In \cite{chandre} we showed  that starting with the dynamical equations for the two-dimensional ionic fluid (with density $n(x,y)$ and velocity field ${\bf v}(x,y)$) and imposing a set of local constraints $\nabla \cdot {\bf v}=0$ and $n=N(\phi)$ around equilibrium, where $\phi$ is the streamfunction, the associated Dirac bracket is written as
\beq
\{F,G\}_*=\int \!d^2x\,  (\lap \phi-N(\phi))[(\lap-\hN)^{-1}F_\phi,(\lap-\hN)^{-1}G_\phi],
\label{eq:PB_D}
\eeq
where $\hN$ is the Fr\'echet derivative of the pseudo-differential function $N$. Here, the canonical Poisson bracket   $[f,g]:=\hz\cdot \nabla f\times \nabla g$.  
Then, the Hamiltonian is written as 
\beq
H=\frac{1}{2}\int\! d^2x \, \left( |\nabla \phi|^2+(N(\phi)+\lambda)^2\right),
\label{eq:H_D}
\eeq
where $\lambda=\lambda(x)$ characterizes the equilibrium and the integration is over  the two-dimensional  cylinder 
${\mathbb R}\times [0,2 \pi]$. 
Compared to the HM  equation,  where the second constraint was $N(\phi)=\phi-\lambda$, we impose here the following  modified constraint:
\beq
N(\phi)=\phi-\frac{1}{2\pi}\int_0^{2 \pi} \phi dy  -\lambda. 
\eeq
Therefore, $\hN=1-P$, with 
%\subsection{Hamitonian and Poisson bracket}
the operator  $P$ defined by 
$$
P(\chi)=\frac{1}{2 \pi}\int_0^{2 \pi} \chi dy\ \,, 
$$ 
which evidently is symmetric in the sense that $\int f P g=\int g P f$. 

The Poisson bracket for the  mHM equation is given by
\begin{equation}
\label{eq:PB_MHMi}
\{F,G\}_*=\int\! d^2x \, (\lap \phi-(1-P)\phi+\lambda)\big[(\lap-1+P)^{-1}F_\phi,(\lap-1+P)^{-1}G_\phi\big],
\end{equation}
and its Hamiltonian is 
$$
H=\frac{1}{2}\int d^2x \left( |\nabla \phi|^2+(\phi-P\phi)^2\right).
$$
To see that this defines the correct Hamiltonian structure,  note that  $H_\phi=-(\lap-1+P) \phi$ and  the mHM equation for $\phi$ is given by $\dot{\phi}=\{\phi,H\}$ which reads as
$$
(\lap-1+P)\dot{\phi}=\big[\lap\phi-\phi+P\phi+\lambda,\phi\big]\,.
$$

In order to obtain the two dynamical equations for the mHM system, we perform the change of variables $\phi \mapsto (\bp,\tp)$ defined by $\phi=\bp+\tp$ and $\bp=P\phi$.  Any observable $F(\phi)=\hat{F}(\bp,\tp)$ has the functional derivative chain rule relation 
$$
F_\phi=(1-P)\hat{F}_\bp+P\hat{F}_\tp,
$$
where we notice that the   part $P\hat{F}_\bp-P\hat{F}_\tp$ is only a function of $x$. 
In what follows we drop the hats on $F$ for simplicity. The Poisson bracket in these new field variables is 
\ben
\{F,G\}_*&=&-\int \!dx \, \lap^{-1}(P F_\bp-PF_\tp)P[\lap\tp-\tp,(\lap-1+P)^{-1}G_\tp]\nonumber\\
&& +\int \!dx \, \lap^{-1}(PG_\bp-PG_\tp)P[\lap\tp-\tp,(\lap-1+P)^{-1}F_\tp]\nonumber\\
&& +\int \!dx \, (\lap \bp+\lambda)P[(\lap-1+P)^{-1}F_\tp,(\lap-1+P)^{-1}G_\tp]\nonumber\\
&&+ \int \!d^2x\, (\lap \tp-\tp)[(\lap-1+P)^{-1}F_\tp,(\lap-1+P)^{-1}G_\tp],
\label{eq:PB_MHM}
\een
and the Hamiltonian is 
\beq
H=\frac{1}{2}\int \!d^2x \, \left(|\nabla \tp|^2+\tp^2+|\nabla \bp|^2 \right),
\label{eq:H_MHM}
\eeq
where we have used the fact that $\int d^2x \nabla \tp \cdot \nabla \bp=0$.
The two dynamical equations for the mHM system are 
\benn
&& \dot{\bp}=\{\bp,H\}=\lap^{-1}P[\lap\tp,\tp],\\
&& \dot{\tp}=\{\tp,H\}=(\lap-1)^{-1}\left( (1-P)[\lap\tp,\tp]+[\lap\tp-\tp,\bp]+[\lap\bp+\lambda,\tp] \right),
\eenn
where we have used the fact that $H_\bp-PH_\tp=-\lap\bp$ and $H_\tp=-\lap\tp+\tp$. 

%\subsection{Casimir invariants}

The Casimir invariants of the Poisson bracket given by Eq.~(\ref{eq:PB_D}) are given by
$$
C(\phi)=\int\! d^2x\, \alpha(\lap\phi -N(\phi)),
$$
where $\alpha$ is scalar function of one variable. For the Poisson bracket given by Eq.~(\ref{eq:PB_MHM}), the condition which determines the Casimir invariants  is given by
$$
C_\bp-PC_\tp+C_\tp=(\lap-1+P) \, \alpha'(\lap\tp-\tp+\lap\bp+\lambda)\,, 
$$
whence we obtain
$$
C(\tp,\bp)=\int \!d^2x \, \alpha(\Delta \tp- \tp+\lap\bp+\lambda)+\int \!dx\,  \beta(x)P\tp,
$$
where $\alpha$ and $\beta$ are two arbitrary scalar functions of one variable. 
Notice in particular that $P\tp(x,y)$ is a local Casimir invariant, obtained for $\alpha=0$ and $\beta(x)=\delta(x-x')$. We have chosen here to restrict to $P\tp=0$.

%\appendix

%\section*{Proof of Jacobi identity}

It should be pointed out that Dirac's theory ensures the Jacobi identity for the bracket~(\ref{eq:PB_MHMi}) before the change of variables. Therefore if the change of variables is invertible, i.e., under the condition $P\tp=0$, then the Jacobi identity is ensured too. However we show explicitly that the Poisson bracket of (\ref{eq:PB_MHM}) satisfies the Jacobi identity unconditionally, i.e., for all field variables $(\bp,\tp)$, not necessarily restricted to $P\tp=0$. In order to demonstrate this, we write the bracket~(\ref{eq:PB_MHM}) as
$$
\{F,G\}_*=\int \!d^2x \,\big((A-P)\tp +\Delta \bp+\lambda\big)\big[A^{-1}((1-P)F_\tp+PF_\bp),A^{-1}((1-P)G_\tp+PG_\bp)\big],
$$
where $A=\Delta-1+P$. In the following, we let $f:=(1-P)F_\tp+PF_\bp$. As was shown in \cite{morr82}, only the functional derivatives of $\{F,G\}$  that take into account the explicit dependence of the bracket on the variables are needed.  These are 
\begin{eqnarray}
&& \{F,G\}_\tp=(A-P)\big[A^{-1}f,A^{-1}g\big],\\
&& \{F,G\}_\bp=P\Delta \big[A^{-1}f,A^{-1}g\big].
\end{eqnarray}
The computation of $\{\{F,G\},H\}$ leads to 
\begin{eqnarray}
\{\{F,G\},H\}&=&\int\! d^2x \,\big((A-P)\tp+\Delta\bp+\lambda\big)
\nonumber\\
&& \hspace{.5 cm}\times \Big[A^{-1}\left((1-P)(A-P)[A^{-1}f,A^{-1}g]+P\Delta[A^{-1}f,A^{-1}g]\right),A^{-1}h\Big].
\label{expression}
\end{eqnarray}
Since $(1-P)(A-P)+P\Delta=A$, Eq.~(\ref{expression}) becomes
$$
\{\{F,G\},H\}=\int d^2x ((A-P)\tp+\Delta\bp+\lambda)\left[[A^{-1}f,A^{-1}g],A^{-1}h]\right. 
$$
Therefore, the Jacobi identity for the bracket $\{\cdot,\cdot\}$ follows from the Jacobi identity for the bracket $[\cdot,\cdot]$. 

 Given the Hamiltonian structure we are now set to use Hamiltonian techniques for instance for investigating equilibrium and stability (e.g., \cite{morrison98,Hol85,amp2a}). In addition, the proposed construction is sufficiently general that it could be used beyond mHM, in more general Hamiltonian zonal flow models.

\end{document}